\documentclass[11pt]{article}
\usepackage{amssymb}
\usepackage{graphicx}
\usepackage{amsfonts}
\usepackage{amsmath}
\usepackage{longtable,lscape}
\usepackage{amscd}
\usepackage{mathrsfs}
\usepackage{latexsym}
\usepackage{bbm}
\usepackage{float}
\usepackage{url}
\usepackage{natbib}

\newcommand{\captionfonts}{\footnotesize}
\makeatletter  
\long\def\@makecaption#1#2{%
  \vskip\abovecaptionskip
  \sbox\@tempboxa{{\captionfonts #1: #2}}%
  \ifdim \wd\@tempboxa >\hsize
    {\captionfonts #1: #2\par}
  \else
    \hbox to\hsize{\hfil\box\@tempboxa\hfil}%
  \fi
  \vskip\belowcaptionskip}
\makeatother 

\title{{\bf Testing Ambiguity and Machina Preferences Within a Quantum-theoretic Framework for Decision-making}}
\author{Diederik Aerts\footnote{Center Leo Apostel for Interdisciplinary Studies and Department of Mathematics, Brussels Free University, Krijgskundestraat 33, 1160 Brussels (Belgium). Email address: \emph{diraerts@vub.ac.be}}
\, 
Suzette Geriente\footnote{Block 28 Lot 29 Phase III F1, Kaunlaran Village, Caloocan City (The Philippines). Email address: \emph{sgeriente83@yahoo.com}}
\,
Catarina Moreira\footnote{Instituto Superior T\'ecnico, University of Lisbon, INESC-ID, Rua Alves Redol 9, 1000-029 Lisbon (Portugal). Email address: \emph{catarina.p.moreira@tecnico.ulisboa.pt}}
\, 
Sandro Sozzo\footnote{School of Business and Research Centre IQSCS, University Road, LE1 7RH Leicester (United Kingdom). Email address: \emph{ss831@le.ac.uk}}}

\date{}

\begin{document}

\maketitle

\begin{abstract}
\noindent
The Machina thought experiments pose to major non-expected utility models challenges that are similar to those posed by the Ellsberg thought experiments to subjective expected utility theory (SEUT). We test human choices in the `Ellsberg three-color example', confirming typical ambiguity aversion patterns, and the `Machina 50/51 and reflection examples', partially confirming the preferences hypothesized by Machina. Then, we show that a quantum-theoretic framework for decision-making under uncertainty recently elaborated by some of us allows faithful modeling of all data on the Ellsberg and Machina paradox situations. In the quantum-theoretic framework subjective probabilities are represented by quantum probabilities, while quantum state transformations enable representations of ambiguity aversion and subjective attitudes toward it.
\end{abstract}

\section{Introduction\label{intro}}
Probability theorists distinguish between probabilities that are known or knowable in principle (`objective probabilities') and probabilities that are not known or knowable. For this reason, the term `risk' is generally used to designate uncertainty situations that can be described by objective probabilities, while the term `ambiguity' to designate uncertainty situations that cannot be described by objective probabilities \cite{k1921}. The so-called `Bayesian paradigm' minimizes this distinction by introducing the notion of `subjective probability': when probabilities are not known, people form their own `beliefs' or `priors' \cite{gps2008}. Savage's `subjective expected utility theory' (SEUT), then, extends `objective EUT' \cite{vnm1944} within the Bayesian paradigm \cite{s1954}. Savage identified a set of axioms allowing to uniquely represent human preferences by expected utility maximization for a unique subjective probability measure satisfying Kolmogorov's axioms of probability theory (`Kolmogorovian probability') \cite{k1933}. 

In a seminal 1961 paper, Daniel Ellsberg presented a set of examples which show that human decisions violate one of the axioms of SEUT, namely, the `sure-thing principle' \cite{e1961}. More specifically, people generally prefer to bet on events with objective probability, rather than on events with subjective probability. And, several decision-making experiments have confirmed this phenomenon known since Ellsberg as `ambiguity aversion' \cite{cw1992}, \cite{ms2014}. 

The `Ellsberg paradox' led many scholars to rethink the foundations of SEUT and to formulate more general decision-making models which could represent subjective probabilities, ambiguity and subjective attitudes toward it. In this respect, the major `non-expected utility models' include, but are not limited to, `Choquet expected utility' \cite{s1989}, `cumulative prospect theory' \cite{tk1992}, `maxmin expected utility' \cite{gs2008}, `$\alpha$-maxmin expected utility' \cite{gmm1995}, `variational preferences' \cite{mmr2006} and `smooth ambiguity preferences' \cite{kmm2005} (see, e.g., \cite{gm2013} and \cite{ms2014} for extensive reviews). 

More recently, Mark Machina presented two Ellsberg-like examples, the `50/51 example' and the `reflection example', which pose to the above non-expected utility models difficulties that are similar to the difficulties posed by the Ellsberg paradox to SEUT \cite{m2009}, \cite{blhp2011}. In particular, the `Machina paradox' violates the `tail-separability property' of Choquet expected utility. The reflection example was tested in a decision-making experiment \cite{lhp2010}, confirming `Machina preferences' against tail-separability, while the 50/51 example has not been tested yet, at the best of our knowledge (Section \ref{SEUT}).

We took a different direction to decision-making under uncertainty, which rests on a successful implementation of the mathematical formalism of quantum theory, in particular, quantum probability, to model cognitive phenomena that have resisted traditional modeling in terms of Kolmogorovian probability, e.g., `conjunctive and disjunctive fallacies', `question order effects', `over- and under-extension in membership and typicality judgments' and `disjunction effect' \cite{a2009}, \cite{abgs2013}, \cite{ags2013}, \cite{bpft2011}, \cite{bb2012}, \cite{hk2013}, \cite{pb2013}, \cite{pnas2014}. According to this approach, in any decision process a contextual interaction occurs  between the decision-maker and the cognitive situation (the `decision-making (DM) entity') that is the object itself of the decision. As a consequence of this contextual interaction, the state of the DM entity may change. This led us to develop a quantum-based state-dependent theoretic framework, able to model human preferences in the Ellsberg and Machina paradox situations, and to represent concrete human decisions in the three-color Ellsberg experiment and the Machina reflection experiment \cite{ast2014}, \cite{as2016}, \cite{ahs2017} (Section \ref{QEUT}).

In the present paper, we proceed further in this direction and report the results of various decision-making experiments that we performed in situations where ambiguity is present. We firstly performed the `Ellsberg three-color experiment', confirming the typical ambiguity aversion pattern that is reported in the experimental literature. Next, we performed the Machina 50/51 and reflection experiments. While the 50/51 experiment confirms the preferences hypothesized by Machina, showing an inversion of preferences incompatible with the predictions of non-expected utility models, the reflection example does not confirm the findings in \cite{lhp2010} and identifies a pattern of preferences that partially agree with existing theoretical proposals (Section \ref{experiments}).

Then, we show that the experimental data in Section \ref{experiments} can be faithfully represented within the above mentioned quantum-theoretic framework in which subjective probabilities are represented by quantum probabilities, while the state of the DM entity and its transformations in a decision-making process allow to represent ambiguity and individual preferences toward it (Section \ref{QEUT}). These results are compatible with some recent approaches that use the mathematical formalism of quantum theory to specifically model decision-making processes, namely, La Mura's modeling of the Allais \cite{a1953} and Ellsberg paradoxes \cite{lm2009}, and Khrennikov's extension of the Aumann theorem \cite{k2015}.

We conclude the paper with some general considerations which agree with the conclusions in \cite{m2009} on the issues of `event-separability properties', like those stated by the sure-thing principle and tail-separability, in concrete human decisions (Section \ref{conclusions}).

\section{Expected utility theory, Ellsberg and Machina paradoxes\label{SEUT}}
The basic mathematical framework of SEUT and its major extensions requires a set ${\mathscr S}$ of (physical) `states of nature', a $\sigma$-algebra ${\mathscr A}\subseteq {\mathscr P}({\mathscr S})$ of subsets of ${\mathscr S}$, called `events', and a subjective probability measure $p:{\mathscr A}\subseteq {\mathscr P}({\mathscr S})\longrightarrow [0,1]$ over $\mathscr A$. Let ${\mathscr X}$ be the set of all `consequences'. An `act' is a function $f: {\mathscr S} \longrightarrow {\mathscr X}$ mapping states into consequences. Next, we introduce a `weak preference relation' (reflexive, symmetric and transitive) $\succsim$ over the Cartesian product $\mathscr{F} \times \mathscr{F}$, where $\mathscr{F}$ is the set of all acts and $\succ$ and $\sim$  respectively denote `strong preference' and `indifference'. Let finally $u: {\mathscr X} \longrightarrow \Re$ be a strictly increasing and continuous `utility function' mapping consequences into real numbers and expressing the decision-maker's taste. 

For the sake of simplicity, we assume that the set ${\mathscr S}$ is discrete and finite, and ${\mathscr X}$ consists of monetary payoffs. Let $\{ E_{1}, E_{2}, \ldots, E_{n}  \}$ be a set of mutually exclusive and exhaustive elementary events, forming a partition of $\mathscr S$. And, let $f$ be the act that, for every  $i \in \{1,2,\ldots,n\}$, associates the event $E_{i}$ with the consequence $x_{i} \in \Re$, so that $f$ can be written as $f=(E_1,x_1;E_2,x_2;\ldots;E_n,x_n)$. Under SEUT, $f$ can be represented as the `expected utility functional' $W(f)=W(E_1,x_1;E_2,x_2;\ldots;E_n,x_n)=\sum_{i=1}^{n} p_i u(x_i)$ where, for every $i \in \{1,2,\ldots,n\}$, $p_i=p(E_i)$ is interpreted as the `subjective probability' that the event $E_i$ occurs, and expresses the decision-maker's beliefs.

Let now $f$ and $g$ be two acts, and let $W(f)$ and $W(g)$ be the corresponding expected utilities. Then, Savage proved that, if suitable axioms are satisfied, including the sure-thing principle, one has $f \succsim g$ if and only if $W(f)\ge W(g)$ \cite{s1954}. We stress that the utility function $u$ is unique (up to positive affine transformations), and the subjective probability distribution $p$, which is defined over a single $\sigma$-algebra ${\mathscr A}$ of events, is unique and satisfies the axioms of Kolmogorovian probability theory \cite{k1933}.

In 1961, Daniel Ellsberg proved in a series of thought experiments that decision-makers generally prefer acts with known (or objective) probabilities to acts with unknown (or subjective) probabilities. Let us, for example, consider the `Ellsberg three-color example', that is, one urn with 30 red balls and 60 balls that are either yellow or black in unknown proportion. One ball will be drawn at random from the urn. Then, free of charge, a person is asked to bet on one of the acts $f_1$, $f_2$, $f_3$ and $f_4$ in Table 1. 
\noindent 
\begin{table} \label{table01}
\begin{center}
\begin{tabular}{|p{1.5cm}|p{1.5cm}|p{1.5cm}|p{1.5cm}|}
\hline
\multicolumn{1}{|c|}{} & \multicolumn{1}{c|}{1/3} & \multicolumn{2}{c|}{2/3} \\
\hline
Act & Red & Yellow & Black \\ 
\hline
\hline
$f_1$ & \$100 & \$0 & \$0 \\ 
\hline
$f_2$ & \$0 & \$0 & \$100 \\ 
\hline
$f_3$ & \$100 & \$100 & \$0 \\ 
\hline
$f_4$ & \$0 & \$100 & \$100 \\ 
\hline
\end{tabular}
\end{center}
{\bf Table 1.} The payoff matrix for the Ellsberg three-color thought experiment.
\end{table}
\noindent 
Ellsberg suggested that, when asked to rank these acts, most people will prefer  $f_1$ over $f_2$ and $f_4$ over $f_3$. Indeed, acts $f_1$ and $f_4$ are `unambiguous', because they are associated with events over known probabilities, and hence provide `clear information'. On the contrary, acts $f_2$ and $f_3$ are `ambiguous', because they are associated with events over unknown probabilities. This attitude of decision-makers is known as `ambiguity aversion' \cite{e1961}. Several experiments have confirmed the `Ellsberg preferences' $f_1 \succ f_2$ and $f_4 \succ f_3$, hence ambiguity aversion (see, e.g., \cite{ms2014}), and only Slovic and Tversky (1974) found `ambiguity attraction'.

Preferences of decision-makers who are sensitive to ambiguity cannot be explained within SEUT, because they violate the sure-thing principle, according to which, preferences should be independent of the common outcome. In the Ellsberg three-color urn, preferences should not depend on whether the common event ``a yellow ball is drawn'' pays off \$0 or \$100. More concretely, SEUT predicts `consistency of decision-makers' preferences', that is, $f_1 \succsim f_2$ if and only if $f_3 \succsim f_4$. A simple calculation shows indeed that it is impossible to assign subjective probabilities  $p_{R}=\frac{1}{3}$, $p_{Y}$ and $p_{B}=\frac{2}{3}-p_{Y}$ such that $W(f_1)>W(f_2)$ and $W(f_4)>W(f_3)$.

Several extensions of SEUT have been put forward, mainly in an axiomatic form, which cope with the `Ellsberg paradox', replacing the sure-thing principle by weaker axioms and representing subjective probabilities by more general, possibly non-Kolmogorovian, mathematical structures. Major proposals include `Choquet expected utility' \cite{s1989}, `cumulative prospect theory' \cite{tk1992}, `maxmin expected utility' \cite{gs2008}, `$\alpha$-maxmin expected utility' \cite{gmm1995}, `variational preferences' \cite{mmr2006} and `smooth ambiguity preferences' \cite{kmm2005} (see, e.g., the reviews in \cite{gm2013} and \cite{ms2014}). 

In 2009, Mark Machina presented two thought experiments, the `50/51 example' and the `reflection example', which challenge the non-expected utility models above in a similar way as the Ellsberg paradox challenges SEUT \cite{m2009},\cite{blhp2011}.

Let us consider the Machina 50/51 example and formulate it in a way that is suitable for our purposes. One urn contains 50 balls that are either red or yellow in unknown proportion and 51 balls that are either black or green in unknown proportion. One ball will be drawn at random from the urn. Then, free of charge, a person is asked to bet on one of the acts $f_1$, $f_2$, $f_3$ and $f_4$ in Table 2. 
\noindent 
\begin{table} \label{table03}
\begin{center}
\begin{tabular}{|p{1.5cm}|p{1.5cm}|p{1.5cm}|p{1.5cm}|p{1.5cm}|}
\hline
\multicolumn{1}{|c|}{} & \multicolumn{2}{c|}{50/101} & \multicolumn{2}{c|}{51/101} \\
\hline
Act & Red & Yellow & Black & Green \\ 
\hline
\hline
$f_1$ & \$202 & \$202 & \$101 & \$101 \\ 
\hline
$f_2$ & \$202 & \$101 & \$202 & \$101 \\ 
\hline
$f_3$ & \$303 & \$202 & \$101 & \$0 \\ 
\hline
$f_4$ & \$303 & \$101 & \$202 & \$0 \\ 
\hline
\end{tabular}
\end{center}
{\bf Table 2.} The payoff matrix for the Machina 50/51 example.
\end{table}
\noindent 
For every $i \in \{R,Y, B,G\}$, let $E_i$ be the elementary event ``a ball of color $i$ is drawn from the urn''. Clearly, all $E_i$s are ambiguous events, while the event ``a red or yellow ball is drawn'' has an objective probability $\frac{50}{101}$ and the event ``a black or green ball is drawn'' has an objective probability $\frac{51}{101}$. As such, the act $f_1$ in Table 2 is unambiguous, while the acts $f_2$, $f_3$ and $f_4$ in the same table are ambiguous. On the other hand, the acts $f_2$ and $f_4$ benefit from the Bayesian advantage offered by the 51st ball yielding 202. Hence, an ambiguity averse decision-maker may prefer $f_1$ over $f_2$ and may be indifferent between $f_3$ and $f_4$. On the contrary, an expected utility maximizer assuming a uniform distribution over the balls will prefer $f_2$ and $f_4$. As argued in \cite{m2009} and \cite{blhp2011}, the `informational advantage' of $f_1$ can more than offset its Bayesian advantage with respect to $f_2$, while $f_3$ does not offer any clear informational advantage with respect to $f_4$. This leads to the `Machina  preferences' $f_1 \succ f_2$ and $f_4 \succ f_3$. \cite{m2009} and \cite{blhp2011} showed that none of the non-expected utility models above can reproduce this pattern, because they all predict $f_1 \succ f_2$ if and only if $f_3 \succ f_4$. In other words, none of these models above can represent preferences revealing this trade-off  between ambiguity aversion and Bayesian advantage.

Let us come to the Machina reflection example `with lower tail shifts'.  Consider one urn with 20 balls, 10 are either red or yellow in unknown proportion, 10 are either black or green in unknown proportion. One ball will be drawn at random from the urn. Then, free of charge, a person is asked to bet on one of the acts $f_1$, $f_2$, $f_3$ and $f_4$ in Table 3.
\noindent
\begin{table} \label{table04}
\begin{center}
\begin{tabular}{|p{1.5cm}|p{1.5cm}|p{1.5cm}|p{1.5cm}|p{1.5cm}|}
\hline
\multicolumn{1}{|c|}{} & \multicolumn{2}{c|}{1/2} & \multicolumn{2}{c|}{1/2} \\
\hline
Act & Red & Yellow & Black & Green \\ 
\hline
\hline
$f_1$ & \$0 & \$50 & \$25 & \$25 \\ 
\hline
$f_2$ & \$0 & \$25 & \$50 & \$25 \\ 
\hline
$f_3$ & \$25 & \$50 & \$25 & \$0 \\ 
\hline
$f_4$ & \$25 & \$25 & \$50 & \$0 \\ 
\hline
\end{tabular}
\end{center}
{\bf Table 3.} The payoff matrix for the Machina reflection example with lower tail shifts.
\end{table}
\noindent
In this case, all the events ``a ball of color $i$ is drawn from the urn'', $i \in \{R, Y, B, G\}$, are ambiguous. In this respect, Machina introduced the notion of `informational symmetry', namely, the events ``the drawn ball is red or yellow'' and ``the drawn ball is black or green'' have known probability equal to $\frac{1}{2}$ and, further, the ambiguity about the distribution of colors is similar in the two events. A decision-maker who is sensitive to informational symmetry, will then prefer act $f_1$ over act $f_2$ and act $f_4$ over act $f_3$, or act $f_2$ over act $f_1$ and act $f_3$ over act $f_4$, while an expected utility maximizer assuming a uniform distribution over the balls will be indifferent between the four acts, as they have the same expected utility.  \cite{m2009} and \cite{blhp2011} showed that, according to the non-expected utility models above, $f_1 \succsim f_2$ if and only if $f_3 \succsim f_4$ and a decision-maker who is sensitive to informational symmetry should exhibit $f_1 \sim f_2$ and $f_3 \sim f_4$.

Let us finally consider the Machina reflection example `with upper tail shifts'. One urn contains 20 balls, 10 are either red or yellow in unknown proportion, 10 are either black or green in unknown proportion. One ball will be drawn at random from the urn. Then, free of charge, a person is asked to bet on one of the acts $f_1$, $f_2$, $f_3$ and $f_4$ defined in Table 4. 
\noindent 
\begin{table} \label{table05}
\begin{center}
\begin{tabular}{|p{1.5cm}|p{1.5cm}|p{1.5cm}|p{1.5cm}|p{1.5cm}|}
\hline
\multicolumn{1}{|c|}{} & \multicolumn{2}{c|}{1/2} & \multicolumn{2}{c|}{1/2} \\
\hline
Act & Red & Yellow & Black & Green \\ 
\hline
\hline
$f_1$ & \$50 & \$50 & \$25 & \$75 \\ 
\hline
$f_2$ & \$50 & \$25 & \$50 & \$75 \\ 
\hline
$f_3$ & \$75 & \$50 & \$25 & \$50 \\ 
\hline
$f_4$ & \$75 & \$25 & \$50 & \$50 \\ 
\hline
\end{tabular}
\end{center}
{\bf Table 4.} The payoff matrix for the Machina reflection example with upper tail shifts.
\end{table}
\noindent 
According to Machina's informational symmetry, people should again prefer act $f_1$ over act $f_2$ (act $f_2$ over act $f_1$) and act $f_4$ over act $f_3$ (act $f_3$ over act $f_4$). On the other hand, the non-expected utility models above predict that $f_1 \succsim f_2$ if and only if $f_3 \succsim f_4$, and a decision-maker who is sensitive to informational symmetry should exhibit $f_1 \sim f_2$ and $f_3 \sim f_4$\cite{m2009}, \cite{blhp2011}.

A decision-making experiment performed by L'Haridon and Placido (2010) confirmed the `Machina preferences' $f_1 \succ f_2$ and  $f_4 \succ f_3$, consistently with an informational symmetry scenario. These authors found that the `tail-separability property' is specifically violated in non-expected utility models with rank-dependent utility, like Choquet expected utility and cumulative prospect theory.

The immediate consequence is that representing Machina preferences needs a completely new theoretical approach to ambiguity and subjective attitudes toward it \cite{lhp2010}.

\section{Analyzing the experiments\label{experiments}}
We investigated the Ellsberg three-color, the 50/51 and the reflection examples collecting data in a decision-making experiment on human participants. The present section is devoted to analyze the results and draw relevant conclusions.

We asked 200 people, chosen among colleagues and friends, to fill a questionnaire in which they had to rank acts according to Tables 1--4 in a within subjects experimental setup. People had on average a basic knowledge of probability theory, but no training in decision theory.\footnote{The experimental study was carried out in accordance with the recommendations of the 'University of Leicester Code of Practice and Research Code of Conduct, Research Ethics Committee of the School of Management' with written informed consent from all subjects. All subjects gave written informed consent in accordance with the Declaration of Helsinki.}

In the first experiment, we tested the Ellsberg three-color example. Participants were provided with a paper similar to the one in Figure 1, in which they had to choose between acts $f_1$ and $f_2$ and, then, between acts $f_3$ and $f_4$ in Table 1, Section \ref{SEUT}. For the sake of simplicity, we assumed that each choice concerned two alternatives, hence indifference between acts was not a possible option.
\begin{figure}
\begin{center}
\includegraphics[scale=0.6]{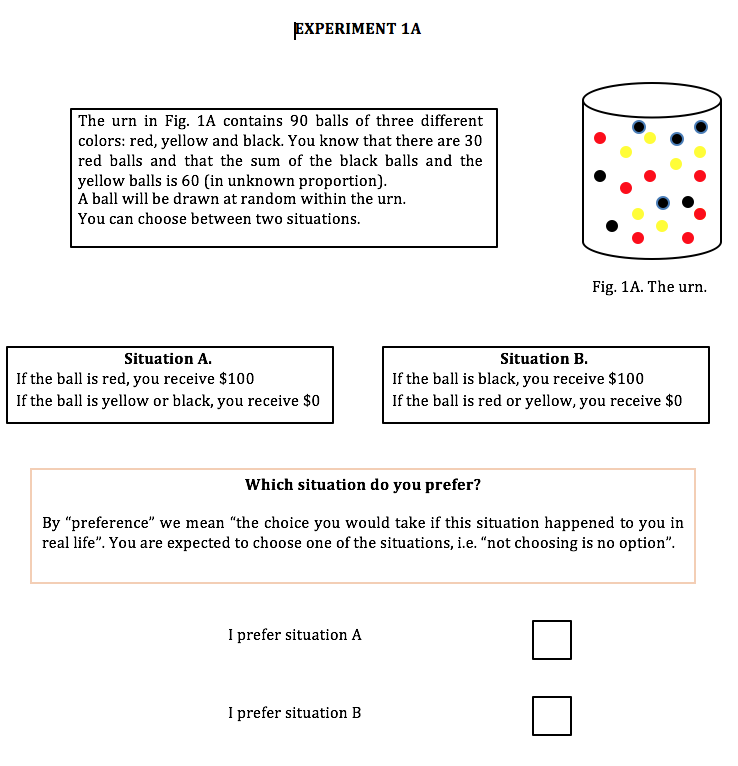}
\end{center}
{\bf Figure 1.} A sample of the questionnaire related to the decision-making experiment on the Ellsberg three-color urn: choice between acts $f_1$ and $f_2$ in Table 1.
\end{figure}
Overall, 125 participants preferred acts $f_1$ and $f_4$, 38 preferred acts $f_1$ and $f_3$, 6 preferred acts $f_2$ and $f_3$, and 31 preferred acts $f_2$ and $f_4$, as summarized in Table 5. This means that 163 participants over 200 preferred act $f_1$ over act $f_2$. This difference is significant (p-value=$1.25 \cdot 10^{-23}$) and entails a `preference weight' of 0.815. Further, 156 participants over 200 preferred act $f_4$ over act $f_3$. This difference is also significant (p-value=$5.48 \cdot 10^{-18}$) and entails a `preference weight' of 0.780. The inversion rate is 0.655. Finally, a significant difference in the choices $f_1f_2$ and $f_3f_4$ was observed  (p-value=$1.91 \cdot 10^{-35}$). This pattern agrees with the Ellsberg preferences and points toward ambiguity aversion, while it cannot be reproduced within SEUT. In addition, these results confirm empirical findings existing in the literature (see, e.g., \cite{ms2014}).
\noindent 
\begin{table} \label{table06}
\begin{center}
\begin{tabular}{|p{5.5cm}|p{1.5cm}|p{1.5cm}|p{1.5cm}|p{1.5cm}|}
\hline

Experiment (N=200) & $f_1f_4$ & $f_1f_3$  & $f_2f_3$ & $f_2f_4$ \\ 
\hline
Three-color example & 125 & 38 & 6 & 31 \\ 
50/51 example & 59 & 57 & 17 & 67 \\ 
Reflection example lower t. s. & 64 & 51 & 59 & 26 \\ 
Reflection example upper t. s. & 80 & 54 & 50 & 16 \\ 
\hline
\end{tabular}
\end{center}
{\bf Table 5.} Number of participants who preferred acts $f_i$ and $f_j$, $i\in \{1,2\}$, $j\in\{3,4\}$, in each experiment.
\end{table}

In the second experiment, we tested the Machina 50/51 example. Participants were provided with a paper in which they had to choose between acts $f_1$ and $f_2$ and between acts $f_3$ and $f_4$ in Table 2, Section \ref{SEUT}. Overall, 59 participants preferred acts $f_1$ and $f_4$, 57 preferred acts $f_1$ and $f_3$, 17 preferred acts $f_2$ and $f_3$, and 67 preferred acts $f_2$ and $f_4$, as summarized in Table 5. This means that 116 participants over 200 preferred act $f_1$ over act $f_2$. This difference is significant (p-value=$2.33 \cdot 10^{-2}$) and entails a `preference weight' of 0.580. Further, 126 participants over 200 preferred act $f_4$ over act $f_3$. This difference is also significant (p-value=$1.93 \cdot 10^{-4}$) and entails a preference weight of 0.630. The inversion rate is 0.380. Finally, a significant difference in the choices $f_1f_2$ and $f_3f_4$ was observed  (p-value=$7.48 \cdot 10^{-7}$). This pattern agrees with the Machina preferences in Section \ref{SEUT}, and can be explained within a trade-off between ambiguity aversion and Bayesian advantage, while it cannot be reproduced either within SEUT or its non-expetcted utility extensions.

The situation is more subtle for what concerns the reflection examples, whose experimental findings are reported in the following.

In the third experiment, we tested the Machina reflection example with lower tail shifts. Participants were provided with a paper in which they had to choose between acts $f_1$ and $f_2$ and between acts $f_3$ and $f_4$ in Table 3, Section \ref{SEUT}. Overall, 64 participants preferred acts $f_1$ and $f_4$, 51 preferred acts $f_1$ and $f_3$, 59 preferred acts $f_2$ and $f_3$, and 26 preferred acts $f_2$ and $f_4$, as summarized in Table 5. This means that 115 participants over 200 preferred act $f_1$ over act $f_2$. This difference is significant (p-value=$3.36 \cdot 10^{-2}$) and entails a `preference weight' of 0.575. Further, 120 participants over 200 preferred act $f_3$ over act $f_4$. This difference is not significant (p-value=$1.58 \cdot 10^{-1}$), which can be interpreted by assuming that decision-makers were on average indifferent between $f_3$ and $f_4$. The preference weight is 0.630, in this case. The inversion rate is 0.615. Finally, a not significant difference in the choices $f_1f_2$ and $f_3f_4$ was observed  (p-value=$0.6533$). The result of the Machina reflection experiment with lower tail shifts does not agree with the empirical findings in \cite{lhp2010}, who found agreement with the Machina preferences in Section \ref{SEUT}. Here, we have instead that act $f_1$ is on average preferred over act $f_2$, while the decision-makers are on average indifferent between acts $f_3$ and $f_4$. This behavior shows that decision-makers are not sensible to informational symmetry, while it may be compatible in particular with Choquet expected utility.

In the fourth experiment, we tested the Machina reflection example with upper tail shifts. Participants were provided with a paper in which they had to choose between acts $f_1$ and $f_2$ and between acts $f_3$ and $f_4$ in Table 4, Section \ref{SEUT}. Overall, 80 participants preferred acts $f_1$ and $f_4$, 54 preferred acts $f_1$ and $f_3$, 50 preferred acts $f_2$ and $f_3$, and 16 preferred acts $f_2$ and $f_4$, as summarized in Table 5. This means that 134 participants over 200 preferred act $f_1$ over act $f_2$. This difference is significant (p-value=$7.89 \cdot 10^{-7}$) and entails a `preference weight' of 0.670. Further, 104 participants over 200 preferred act $f_3$ over act $f_4$. This difference is not significant (p-value=$5.73 \cdot 10^{-1}$), which can be interpreted by assuming that decision-makers were on average indifferent between $f_3$ and $f_4$. The preference weight is 0.620, in this case. The inversion rate is 0.650. Finally, a significant difference in the choices $f_1f_2$ and $f_3f_4$ was observed  (p-value=$8.18 \cdot 10^{-3}$). Also in this case, the result of the Machina reflection experiment with upper tail shifts does not agree with the empirical findings in \cite{lhp2010}, who found agreement with the Machina preferences in Section \ref{SEUT}. Here, we have instead that act $f_1$ is on average preferred over act $f_2$, while the decision-makers are on average indifferent between acts $f_3$ and $f_4$. This behavior shows that decision-makers are not sensible to informational symmetry, while it may be compatible in particular with Choquet expected utility.

As we can see, the experimental tests on the Machina reflection examples do not confirm the experimental pattern found in \cite{lhp2010}, while the overall experimental findings cannot be explained within a unique theoretical proposal within the ones sketched in Section \ref{SEUT}.

We will show in the next sections that all the results above can instead be modeled within a quantum-theoretic framework elaborated by some of us \cite{as2016}, \cite{ahs2017}. But, before proceeding further, we need to recall the essentials of such quantum-theoretic framework.

\section{A quantum-theoretic framework to model human decisions\label{QEUT}}
We present in this section a unified framework to model events, states, subjective probabilities, acts, preferences and decisions within the quantum formalism sketched in \ref{quantummathematics}. The quantum-theoretic framework was firstly presented in \cite{as2016} and then developed in \cite{ahs2017}, and is a first step toward the elaboration of `state-dependent EUT' where subjective probabilities are represented by quantum probabilities.

We firstly introduce some basic notions and definitions that are needed to operationally characterize the cognitive aspects of human decisions under uncertainty.

(i) The cognitive situation that is the object of the decision identifies a DM entity in a defined state $p_v$. We denote by $\Sigma_{DM}$ the set of all possible states of the DM entity. This state has a cognitive nature, hence it should be distinguished from a physical state (state of nature). The cognitive state captures mathematical aspects of ambiguity.

(ii) A contextual interaction of a cognitive, {\it not physical}, nature occurs between the decision-maker and the DM entity. This contextual interaction may determine a change of state of the DM entity. 

(iii) The type of state change of the DM entity induced by the contextual interaction with the decision-maker gives information on 
the decision-maker's attitudes toward ambiguity  (ambiguity aversion, ambiguity attraction, etc.).

(iv) Events are related to measurements that can be performed on the DM entity. For each state $p_v$ of the DM entity, each event $E$ is associated with a probability $\mu_{v}(E)$ that $E$ occurs when the DM entity is in the state $p_v$.

(v) The DM entity, its states, events, probabilities and the decision-making process are modeled by using the formalism of quantum theory in \ref{quantummathematics}. In particular, the state of the DM entity identifies a single quantum probability distribution via Born's rule. We interpret this quantum probability distribution, which is generally non-Kolmogorovian, as the subjective probability distribution associated with the specific decision process.

Let us assume, as in Section \ref{SEUT}, that the set ${\mathscr S}$ of states of nature is discrete and finite, and let $\{ E_1, E_2, \ldots, E_n \}$ be a set of mutually exclusive and exhaustive elementary events. Let ${\mathscr X}$ be the set of consequences, which denote monetary outcomes. An act is a function $f:{\mathscr S} \longrightarrow {\mathscr X}$, and ${\mathscr F}$ is the set of all acts. If the act $f$ maps the event $E_i$ into the outcome $x_i\in \Re$, we can write $f=(E_1,x_1;\ldots;E_n,x_n)$. We assume that a utility function $u: {\mathscr X} \longrightarrow \Re$ exists over the set of consequences which incorporates individual preferences toward risk.

We refer to the quantum mathematics in \ref{quantummathematics}. The DM entity is associated with a Hilbert space $\cal H$ over the field $\mathbb C$ of complex numbers. Since $n$ is the number of elementary events, the space $\cal H$ can be chosen to be isomorphic to the Hilbert space ${\mathbb C}^n$ of n-tuples of complex numbers. Let $\{ |\alpha_1\rangle, |\alpha_2\rangle, \ldots, |\alpha_n\rangle\}$ be the canonical orthonormal (ON) basis of ${\mathbb C}^n$, that is, $|\alpha_1\rangle=(1,0,\ldots 0)$, \ldots, $|\alpha_n\rangle=(0,0,\ldots n)$. The event $E_i$ is then represented by the orthogonal projection operator $P_i=|\alpha_i \rangle \langle \alpha_i|$, $i \in \{1,\ldots, n \}$.

For every state $p_v \in \Sigma_{DM}$ of the DM entity, represented by the unit vector $|v\rangle=\sum_{i=1}^{n}\langle \alpha_i|v\rangle |\alpha_i\rangle\in {\mathbb C}^n$, the quantum probability distribution
\begin{equation}
\mu_{v}: P \in {\mathscr L}( {\mathbb C}^n) \longmapsto  \mu_{v}(P)\in [0,1]
\end{equation}
(${\mathscr L}( {\mathbb C}^n)$ is the lattice of all orthogonal projection operators over the complex Hilbert space ${\mathbb C}^n$) induced by Born's rule, represents the subjective probability that the event $E$, represented by the orthogonal projection operator $P$, occurs when the DM entity is in the state $p_v$. Thus, in particular,
\begin{equation} \label{quantumprobability}
\mu_{v}(E_i)=\langle v|P_i|v\rangle=|\langle \alpha_i|v\rangle|^{2}
\end{equation}
for every $i\in \{ 1, \ldots, n\}$.

Suppose that, when the decision-maker is presented with a choice between the acts $f$ and $g$, the DM entity is in the initial state $p_{v_0}$. This state is interpreted as the state of the DM entity when no cognitive context is present. As the decision-maker starts pondering between $f$ and $g$, this mental action can be described as a cognitive context interacting with the DM entity and changing its state. The type of state change directly depends on the decision-maker's attitude toward ambiguity. More precisely, if the DM entity is in the initial state $p_{v_0}$ and the decision-maker is asked to choose between the acts $f$ and $g$, a given attitude toward ambiguity, say ambiguity aversion, will determine a given change of state of the DM entity to a state $p_v$, leading the decision-maker to prefer, say $f$ to $g$. But, a different attitude toward ambiguity will determine a different change of state of the DM entity to a state $p_{w}$, leading the decision-maker to prefer $g$ to $f$. In this way, different attitudes toward ambiguity are formalized by different changes of state inducing different subjective probabilities. 

The considerations above suggest associating the DM entity with a `family of subjective probability distributions', represented by quantum probabilities $\{\mu_{v}:{\mathscr L}( {\mathbb C}^n) \longrightarrow [0,1] \}_{p_{v} \in \Sigma_{DM}}$, and parametrized by the state $p_v$ of the DM entity. In a concrete choice between two acts we will derive the exact state, hence the specific subjective probability distribution that represents the actual choice.

Let us move on to the representation of acts. The act $f=(E_1,x_1;\ldots;E_n,x_n)$ is represented by the Hermitian operator
\begin{equation} \label{quantumact}
\hat{F}=\sum_{i=1}^{n} u(x_i)P_i=\sum_{i=1}^{n} u(x_i)|\alpha_i\rangle\langle \alpha_i|
\end{equation}
For every $p_v \in \Sigma_{DM}$, we introduce the functional `expected utility in the state $p_v$' $W_v:{\mathscr F} \longrightarrow \Re$ as follows. For every $f \in {\mathscr F}$,
\begin{eqnarray} \label{quantumexpectedutility}
W_{v}(f)&=&\langle v| \hat{F}| v \rangle=\langle v| \Big ( \sum_{i=1}^{n} u(x_i)P_i   \Big ) |v\rangle \nonumber \\
=\sum_{i=1}^{n} u(x_i) \langle v|P_i|v\rangle&=&\sum_{i=1}^{n} u(x_i) |\langle \alpha_i|v\rangle|^{2}=\sum_{i=1}^{n}u(x_i) \mu_{v}(E_i)
\end{eqnarray}
because of (\ref{quantumprobability}) and (\ref{quantumact}). Equation (\ref{quantumexpectedutility}) generalizes the expected utility formula of SEUT. We note that the expected utility explicitly depends on the state $p_v$ of the DM entity. This means that, for two acts $f$ and $g$, two states $p_v$ and $p_w$ may exist such that $W_{v}(f)>W_{v}(g)$,  but $W_{w}(f)<W_{w}(g)$, depending on subjective attitudes toward ambiguity. This suggests introducing a state-dependent preference relation $\succsim_{v}$ on the set of acts $\mathscr F$, as follows. 

For every $f,g \in {\mathscr F}$, $p_{v}\in \Sigma_{DM}$,
\begin{equation} \label{statedep}
f \succsim_{v} g \ {\rm iff} \ W_{v}(f) \ge W_{v}(g)
\end{equation}
It follows that the DM entity incorporates the presence of ambiguity, as the quantum probability distribution representing subjective probabilities depends on the state of the DM entity. Furthermore, the way in which the state of the DM entity changes in the interaction with the decision-maker incorporates people's attitude toward ambiguity, as it determines the state-dependent preference relation $\succsim_{v}$. 

The state-dependence enables the `inversion of preferences' observed in the Ellsberg and Machina paradox situations in Section \ref{experiments}, as we will show in the next sections.

\subsection{Application to the Ellsberg paradox\label{quantumellsberg}}
We show in this section that the quantum-theoretic framework presented in Section \ref{QEUT} can be successfully applied to model the data collected on the Ellsberg three-color example in Section \ref{experiments}.

Referring to Section \ref{QEUT}, the DM entity is the urn with 30 red balls and 60 yellow or black balls in unknown proportion, which is assumed to be in a (cognitive) state $p_{v}$. Let $E_R$, $E_Y$ and $E_B$ denote the exhaustive and mutually exclusive elementary events ``a red ball is drawn'', ``a yellow ball is drawn'' and ``a black ball is drawn'', respectively. The events define a `color measurement' on the DM entity with three outcomes, corresponding to the three colors red, yellow and black. The color measurement is represented by a Hermitian operator with eigenvectors $|R\rangle=(1,0,0)$, $|Y\rangle=(0,1,0)$,  and $|B\rangle=(0,0,1)$ or, equivalently, by the spectral family $\{ P_{i}=|i\rangle\langle i | \ | \ i \in \{R,Y,B \}  \}$. In other terms, the event $E_i$ is represented by the orthogonal projection operator $P_i=|i\rangle\langle i|$, $i \in \{R,Y,B\}$. In the canonical basis of ${\mathbb C}^{3}$, a state $p_v$ of the DM entity is instead represented by the unit vector

\begin{equation}
|v \rangle=\sum_{i \in \{R,Y,B\}} \rho_i e^{i \theta_{i}}|i\rangle=(\rho_R e^{i \theta_{R}},\rho_Y e^{i \theta_{Y}},\rho_B e^{i \theta_{B}})
\end{equation}

By using the Born's rule of quantum probability in (\ref{quantumprobability}), the probability $\mu_{v}(E_i)$ of drawing a ball of color $i$, $i\in \{R,Y,B\}$, when the DM entity is in a state $p_v$, is given by

\begin{equation}
\mu_{v}(E_i)=\langle v | P_{i} | v \rangle=|\langle i | v \rangle|^{2}=\rho_{i}^{2}
\end{equation}

We have $\rho_{R}^{2}=\frac{1}{3}$, as the urn contains 30 red balls. Hence, a state $p_v$ of the DM entity is represented by the unit vector 

\begin{equation}  \label{ellsbergstate}
|v \rangle=(\frac{1}{\sqrt{3}} e^{i \theta_{R}},\rho_Y e^{i \theta_{Y}}, \sqrt{\frac{2}{3}-\rho_{Y}^{2}} e^{i \theta_{B}})
\end{equation}

Since the DM entity is a cognitive entity, `cognitive contexts' have an influence on its state, and will generally determine a specific state change. This is what happens in the cognitive realm in analogy with the physical realm, where a physical context will in general change the physical state of a physical entity. Hence, whenever a decision-maker is asked to ponder between the choice of $f_1$ and $f_2$, the pondering itself, before a choice is made, is a cognitive context, and hence it changes in general the state of the DM entity. Similarly, whenever a decision-maker is asked to ponder between the choice of $f_3$ and $f_4$, also this introduces a cognitive context, before the choice is made  -- and a context which in general is different from the one introduced by pondering about the choice between $f_1$ and $f_2$ -- which will in general change the state of the DM entity.

Let us now introduce a state $p_0$ describing the situation where no cognitive context is present. This is the initial state of the DM entity, and symmetry considerations suggest to represent it by the unit vector $|v_0\rangle=\frac{1}{\sqrt{3}}(1,1,1)$. Then, a pondering about the choice between $f_1$ and $f_2$ will make the state $p_0$ of the DM entity change to a state $p_{w_1}$ that is generally different from the state $p_{w_2}$ in which the DM entity changes from $p_0$ when pondering about a choice between $f_3$ and $f_4$.

Let us then come to the representation of the acts $f_1$, $f_2$, $f_3$ and $f_4$ in Table 1, Section \ref{SEUT}. For the utility values $u(0)$ and $u(100)$, the acts $f_1$, $f_2$, $f_3$ and $f_4$ are respectively represented by the Hermitian operators
\begin{eqnarray}
\hat{F}_{1}&=&u(100)P_R+u(0)P_Y+u(0)P_B \label{f1}\\
\hat{F}_{2}&=&u(0)P_R+u(0)P_Y+u(100)P_B \label{f2}\\
\hat{F}_{3}&=&u(100)P_R+u(100)P_Y+u(0)P_B\label{f3} \\
\hat{F}_{4}&=&u(0)P_R+u(100)P_Y+u(100)P_B \label{f4}
\end{eqnarray}
The corresponding expected utilities in a state $p_{v}$ of the DM entity are 
\begin{eqnarray}
W_{v}(f_1)&=&\langle v| \hat{F}_{1}|v\rangle=\frac{1}{3}u(100)+\frac{2}{3}u(0) \label{w1}\\
W_{v}(f_2)&=&\langle v| \hat{F}_{2}|v\rangle=(\frac{1}{3}+\rho_Y^2)u(0)+(\frac{2}{3}-\rho_{Y}^{2})u(100) \label{w2} \\
W_{v}(f_3)&=&\langle v| \hat{F}_{3}|v\rangle=(\frac{1}{3}+\rho_Y^2)u(100)+(\frac{2}{3}-\rho_{Y}^{2})u(0) \label{w3}\\
W_{v}(f_4)&=&\langle v| \hat{F}_{4}|v\rangle=\frac{1}{3}u(0)+\frac{2}{3}u(100) \label{w4}
\end{eqnarray}
We note that $W_{v}(f_1)$ and $W_{v}(f_4)$ do not depend on the state $p_v$, as $f_1$ and $f_4$ are `ambiguity-free acts', while $W_{v}(f_2)$ and $W_{v}(f_3)$ do depend on $p_v$. This means that it is possible to find a state $p_{w_1}$ such that $W_{w_1}(f_1) > W_{w_1}(f_2)$,  and a state $p_{w_2}$ such that $W_{w_2}(f_4) > W_{w_2}(f_3)$. The states $p_{w_1}$ and $p_{w_2}$ reproduce Ellsberg preferences, in agreement with an ambiguity aversion attitude.

We found in the Ellsberg three-color experiment presented in Section \ref{experiments} that the preference weight of participants preferring $f_1$ over $f_2$ was 0.815, and the preference weight of participants preferring $f_4$ over $f_3$ was 0.780. A quantum model for these data can be constructed, for example, by finding two orthogonal states $p_{w_1}$ and $p_{w_2}$, represented by the unit vectors $|w_1\rangle$ and $|w_2\rangle$, respectively, such that
\begin{eqnarray}
\langle w_1|\hat{F}_{1}-\hat{F}_{2}|w_1\rangle&=&0.815 \label{data1} \\
\langle w_2|\hat{F}_{4}-\hat{F}_{3}|w_2\rangle&=&0.780 \label{data2}
\end{eqnarray} 
where $\hat{F}_{i}$, $i\in \{1,2,3,4\}$, are defined in (\ref{f1})--(\ref{f4}). Let us assume, for the sake of simplicity, a risk averse utility function $u(x)=\sqrt{x}$, hence $u(0) = 0$ and $u(100)=10$. Then, in the canonical basis of ${\mathbb C}^{3}$, a solution is
\begin{eqnarray}
|w_1 \rangle& = &(0.577, 0.644, 0.502) \label{w1ellsberg}  \\
|w_2 \rangle& = &(0.577, 0.505 e^{i 238.48^\circ}, 0.641 e^{i 120.46^\circ}) \label{w2ellsberg}
\end{eqnarray}
We however notice that the solution is not unique.

The states $p_{w_1}$ and $p_{w_2}$, represented by the unit vectors in (\ref{w1ellsberg}) and (\ref{w2ellsberg}), identify the subjective probability distributions $\mu_{w_1}$ and $\mu_{w_2}$, respectively, reproducing the ambiguity aversion pattern of the experiment. According to $\mu_{w_1}$, people estimate on average 37.3 yellow balls and 22.7 black balls, while they estimate on average 23 yellow balls and 37 black balls according to $\mu_{w_2}$. 
However, generally speaking, preferences do depend on the state $p_v$ of the DM entity.

This completes the construction of a quantum representation of the Ellsberg three-color experiment in Section \ref{experiments}, which follows the prescriptions in Section \ref{QEUT}.


\subsection{Application to the Machina paradox\label{quantummachina}}
The quantum-theoretic framework in Section \ref{QEUT} also allows straightforward modeling of the Machina 50/51 and reflection examples in Section \ref{experiments}, as we show below.

`Machina 50/51 example'. The DM entity is the urn with 50 red or yellow balls and 51 black or green balls, in both cases in unknown proportion. The DM entity is associated with the Hilbert space $\mathbb{C}^{4}$ over complex numbers. We denote by $(1,0,0,0)$, $(0,1,0,0)$, $(0,0,1,0)$ and $(0,0,0,1)$ the unit vectors of the canonical basis of ${\mathbb C}^{4}$.

Let us now consider the mutually exclusive and exhaustive elementary events $E_R$: ``a red ball is drawn'', $E_Y$: ``a yellow ball is drawn'', $E_B$: ``a black ball is drawn'', and $E_G$: ``a green ball is drawn''. The events $E_{i}$, $i \in \{R,Y,B,G\}$ define a color measurement on the DM entity. This measurement has four outcomes corresponding to the four colors red, yellow, black and green, and it is represented by a Hermitian operator with eigenvectors $|R\rangle=(1,0,0,0)$, $|Y\rangle=(0,1,0,0)$, $|B\rangle=(0,0,1,0)$ and $|G\rangle=(0,0,0,1)$ or, equivalently, by the  spectral family $\{ P_{i}=|i\rangle\langle i | \ | \ i \in  \{R,Y,B,G\} \}$. Thus, the event $E_i$ is represented by the orthogonal projection operator $P_i$, $i\in \{R,Y,B,G\}$. In the canonical basis of ${\mathbb C}^{4}$, a generic state $p_v$ of the Machina entity, which has a cognitive nature, is represented by the unit vector
\begin{equation} \label{general50/51state}
|v \rangle=\sum_{i \in \{R,Y,B,G\}} \rho_i e^{i \theta_{i}}|i\rangle=(\rho_R e^{i \theta_{R}}, \rho_Y e^{i \theta_{Y}}, \rho_B e^{i \theta_{B}},  \rho_G e^{i \theta_{G}})
\end{equation}
By using Born's rule, we can write the probability $\mu_{v}(E_i)$ of drawing a ball of color $i$, when the DM entity is in the state $p_v$ is given by
\begin{equation}
\mu_{v}(E_i)=\langle v | P_{i} | v \rangle=|\langle i | v \rangle|^2=\rho_{i}^{2}
\end{equation}
for every $i \in \{R,Y,B,G\}$. The Machina 50/51 example requires that $\rho_{R}^{2}+\rho_{Y}^{2}=\frac{50}{101}$ and $\rho_{B}^{2}+\rho_{G}^{2}=\frac{51}{101}$, which allows us to write (\ref{general50/51state}) as
\begin{equation} \label{specific50/51state}
|v \rangle=(\rho_R e^{i \theta_{R}}, \sqrt{\frac{50}{101}-\rho_{R}^{2}} e^{i \theta_{Y}}, \rho_B e^{i \theta_{B}},  \sqrt{\frac{51}{101}-\rho_{B}^{2}} e^{i \theta_{G}})
\end{equation}

As in the Ellsberg case, we assume that the initial state $p_0$ of the DM entity completely reflects the initial information about the different colors. Thus, $p_0$ is represented by the unit vector $|v_0\rangle=(\sqrt{\frac{50}{101}},\sqrt{\frac{50}{101}},\sqrt{\frac{51}{101}},\sqrt{\frac{51}{101}})$. Whenever the decision-maker is presented with the Machina 50/51 example, her/his pondering about the choices to make gives rise to a cognitive context, which changes the state of the DM entity from $p_0$ to a generally different state $p_{v}$, represented by the unit vector $|v\rangle$, as in (\ref{general50/51state}). Hence, the pondering about a choice between $f_1$ and $f_2$ will make the initial state $p_0$ of the DM entity change to a state $p_{w_1}$ that is generally different from the state $p_{w_2}$ in which the DM entity changes from the initial state $p_0$ in a pondering about the choice between $f_3$ and $f_4$. 

Let us now represent the acts $f_1$, $f_2$, $f_3$ and $f_4$ in Table 2, Section \ref{SEUT}. For a given utility function $u$, we respectively associate $f_1$, $f_2$, $f_3$ and $f_4$ with the Hermitian operators
\begin{eqnarray}
\hat{F}_{1}&=&u(202)( P_R+P_Y)+u(101) ( P_B+P_G )\label{f1_50/51}\\
\hat{F}_{2}&=&u(202)(P_R+P_B)+u(101)(P_Y+P_G) \label{f2_50/51}\\
\hat{F}_{3}&=&u(303)P_R+u(202)P_Y+u(101)P_B+u(0)P_G \label{f3_50/51} \\
\hat{F}_{4}&=&u(303)P_R+u(101)P_Y+u(202)P_B+u(0)P_G\label{f4_50/51}
\end{eqnarray}
The corresponding expected utilities in a state $p_{v}$ are 
\begin{eqnarray}
W_{v}(f_1)&=&\langle v| \hat{F}_{1}|v\rangle=\frac{50}{101}u(202)+\frac{51}{101}u(101) \label{W1_50/51}\\
W_{v}(f_2)&=&\langle v| \hat{F}_{2}|v\rangle=u(202)\rho_{R}^{2}+u(101)\rho_{y}^{2}+u(202)\rho_{B}^{2}+u(101)\rho_{G}^{2} \label{W2_50/51} \\
W_{v}(f_3)&=&\langle v| \hat{F}_{3}|v\rangle=u(303)\rho_R^{2}+u(202)\rho_B^{2}+u(101)\rho_Y^{2}+u(0)\rho_G^{2} \label{W3_50/51}\\
W_{v}(f_4)&=&\langle v| \hat{F}_{4}|v\rangle=u(303)\rho_R^{2}+u(101)\rho_B^{2}+u(202)\rho_Y^{2}+u(0)\rho_G^{2} \label{W4_50/51}
\end{eqnarray}
We note that the expected utility in (\ref{W1_50/51}) does not depend on the state $p_{v}$, while the expected utilities in (\ref{W2_50/51})--(\ref{W4_50/51}) do depend on $p_{v}$, which expresses the fact that the act $f_1$ is unambiguous, while $f_2$, $f_3$ and $f_4$ are ambiguous. This means that it is in principle possible to find two states $p_{w_1}$ and $p_{w_2}$ of the DM entity such that $W_{w_1}(f_1)>W_{w_1}(f_2)$ and $W_{w_2}(f_4)>W_{w_2}(f_3)$, in agreement with the Machina preferences n Section \ref{SEUT}.

We finally represent the data collected on the Machina 50/51 example in Section \ref{experiments}. We found that the preference weight of participants preferring $f_1$ over $f_2$ was 0.580, while the preference weight of participants preferring $f_4$ over $f_3$ was 0.630. To construct a quantum-theoretic model for these data, we need to find two orthogonal states $p_{w_1}$ and $p_{w_2}$, respectively represented by the unit vectors $|w_1\rangle$ and $|w_2\rangle$ of $\mathbb{C}^{4}$, such that
\begin{eqnarray}
\langle w_1|\hat{F}_{1}-\hat{F}_{2}|w_1\rangle=0.580\label{data1_50/51} \\
\langle w_2|\hat{F}_{4}-\hat{F}_{3}|w_2\rangle=0.630\label{data2_50/51}
\end{eqnarray}
where $\hat{F}_{i}$, $i\in\{1,2,3,4\}$, are defined in (\ref{f1_50/51})--(\ref{f4_50/51}). Assuming again a risk averse utility function $u(x) = \sqrt{x}$, so that $u(202) \approx 14.213$ and $u(101) \approx 10.050$, we have that a solution of (\ref{data1_50/51}) and (\ref{data2_50/51}) is given by
\begin{eqnarray}
|w_1 \rangle&=&(0.487, 0.508, 0.345, 0.621e^{i 90^{\circ}}) \nonumber\\
					&=&(0.487, 0.508, 0.345, 0.62i) \label{w1_50/51}  \\
|w_2 \rangle&=&(0.605 e^{i 90^{\circ}}, 0.359, 0.530 e^{i 180^{\circ}},0.474)\nonumber\\
					&=& (0.605i, 0.359, -0.530,0.474)\label{w2_50/51}
\end{eqnarray}
in the canonical basis of ${\mathbb C}^{4}$. We remind however that such a solution is not unique. 

The states $p_{w_1}$ and $p_{w_2}$ represented in (\ref{w1_50/51}) and (\ref{w2_50/51}) identify the subjective probability distributions $\mu_{w_1}$ and $\mu_{w_2}$, respectively, reproducing the Machina preferences in the 50/51 experiment. Consider also that preferences generally depend on the state $p_v$ of the DM entity. 

This completes the construction of a quantum-theoretic model for the Machina 50/51 experiment, which follows the prescriptions in Section \ref{QEUT}.

`Machina reflection example with lower tail shifts'. The DM entity is the urn with 10 red or yellow balls and 10 black or green balls, in both cases in unknown proportion. A possible (cognitive) state $p_v$ of the DM entity is represented by the unit vector $|v\rangle \in {\mathbb C}^{4}$.

Let us consider again the elementary, exhaustive and mutually exclusive events $E_i$: ``a ball of color $i$ is drawn'', $i \in \{R,Y,B,G\}$. They define a color measurement with four outcomes corresponding to the four colors, and is represented by a Hermitian operator with eigenvectors $|R\rangle=(1,0,0,0)$, $|Y\rangle=(0,1,0,0)$, $|B\rangle=(0,0,1,0)$ and $|G\rangle=(0,0,0,1)$ or, equivalently, by the  spectral family $\{ P_{i}=|i\rangle\langle i | \ | \ i\in \{R,Y,B,G\} \}$. Thus, the event $E_i$ is represented by the orthogonal projection operator $P_i$, $i\in \{R,Y,B,G\}$. In the canonical basis of ${\mathbb C}^{4}$, a state $p_v$ of the DM entity is represented by the unit vector
\begin{equation}
|v \rangle=\sum_{i \in \{R,Y,B,G\}} \rho_i e^{i \theta_{i}}|i\rangle=(\rho_R e^{i \theta_{R}}, \rho_Y e^{i \theta_{Y}}, \rho_B e^{i \theta_{B}},  \rho_G e^{i \theta_{G}})
\end{equation}
The probability $\mu_{v}(E_i)$ of drawing a ball of color $i$, $i\in \{R,Y,B,G\}$, when the Machina entity is in a state $p_v$ is given, through Born's rule, by
\begin{equation}
\mu_{v}(E_i)=\langle v | P_{i} | v \rangle=|\langle i | v \rangle|^2=\rho_{i}^{2}
\end{equation}
The reflection with lower tail shifts situation requires that $\rho_{R}^{2}+\rho_{Y}^2=1/2=\rho_{B}^{2}+\rho_{G}^2$. Hence, a state $p_v$ of the DM entity is represented by the unit vector 
\begin{equation} \label{machinastate}
|v \rangle=(\rho_R e^{i \theta_{R}},\sqrt{\frac{1}{2}-\rho_R^2} e^{i \theta_{Y}}, \rho_B e^{i \theta_{B}}, \sqrt{\frac{1}{2}-\rho_B^2} e^{i \theta_{G}})
\end{equation}

We suppose that the initial state $p_0$ of the DM entity expresses the initial symmetry in the distribution of colors, that is, $p_0$ is represented by the unit vector $|v_0\rangle=\frac{1}{2}(1,1,1,1)$. Whenever the decision-maker is presented with the Machina paradox situation, her/his pondering about the choices to make gives rise to a cognitive context, which changes the state of the DM entity from $p_0$ to a generally different state $p_{w}$, represented by the unit vector $|w\rangle$, as in (\ref{machinastate}). In this framework, the pondering about a choice between $f_1$ and $f_2$ will make the initial state $p_0$ of the DM entity change to a state $p_{w_1}$ that is generally different from the state $p_{w_2}$ in which the DM entity changes from the initial state $p_0$ in a pondering about the choice between $f_3$ and $f_4$.

Let us now represent the acts $f_1$, $f_2$, $f_3$ and $f_4$ in Table 3, Section \ref{SEUT}. For a given utility function $u$, we respectively associate $f_1$, $f_2$, $f_3$ and $f_4$ with the Hermitian operators
\begin{eqnarray}
\hat{F}_{1}&=&u(0)P_R+u(50)P_Y+u(25)P_B+u(25)P_G \label{f1mach}\\
\hat{F}_{2}&=&u(0)P_R+u(25)P_Y+u(50)P_B+u(25)P_G \label{f2mach}\\
\hat{F}_{3}&=&u(25)P_R+u(50)P_Y+u(25)P_B+u(0)P_G \label{f3mach} \\
\hat{F}_{4}&=&u(25)P_R+u(25)P_Y+u(50)P_B+u(0)P_G \label{f4mach}
\end{eqnarray}
The corresponding expected utilities in a state $p_{v}$ are 
\begin{eqnarray}
W_{v}(f_1)&=&\langle v| \hat{F}_{1}|v\rangle=u(0)\rho_R^2+u(50)\rho_Y^2+\frac{1}{2}u(25) \label{w1mach}\\
W_{v}(f_2)&=&\langle v| \hat{F}_{2}|v\rangle=u(0)\rho_R^2+u(25)\rho_Y^2+u(50)\rho_B^2+u(25)\rho_G^2 \label{w2mach} \\
W_{v}(f_3)&=&\langle v| \hat{F}_{2}|v\rangle=u(25)\rho_R^2+u(50)\rho_Y^2+u(25)\rho_B^2+u(0)\rho_G^2 \label{w3mach}\\
W_{v}(f_4)&=&\langle v| \hat{F}_{4}|v\rangle=\frac{1}{2}u(25)+u(50)\rho_B^2+u(0)\rho_G^2 \label{w4mach}
\end{eqnarray}
All expected utilities depend on the state $p_v$ in this case, hence it is possible to find a state $p_{w_1}$ such that $W_{w_1}(f_1) > W_{w_1}(f_2)$, and a state $p_{w_2}$ such that $W_{w_2}(f_3) > W_{w_2}(f_4)$, in agreement with the findings on the reflection example with lower tail shifts (Section \ref{experiments}). We found that the preference weight of participants preferring $f_1$ over $f_2$ was 0.575, and the preference weight of participants preferring $f_3$ over $f_4$ was 0.550. 

A quantum mechanical model for the experimental data above can be constructed by finding two orthogonal states $p_{w_1}$ and $p_{w_2}$, represented by the unit vectors $|w_1\rangle$ and $|w_2\rangle$, respectively, such that
\begin{eqnarray}
\langle w_1|\hat{F}_{1}-\hat{F}_{2}|w_1\rangle=0.575 \label{data1mach} \\
\langle w_2|\hat{F}_{3}-\hat{F}_{4}|w_2\rangle=0.550 \label{data2mach}
\end{eqnarray} 
where $\hat{F}_{i}$, $i\in \{1,2,3,4\}$, are defined in (\ref{f1mach})--(\ref{f4mach}). We assume again the risk averse utility $u(x) = \sqrt{x}$, so that $u(50)=\sqrt{50} \approx 7.071$ and $u(25)=\sqrt{25}=5$. Then, in the canonical basis of ${\mathbb C}^{4}$, a solution is given by
\begin{eqnarray}
|w_1 \rangle&=&(0.333, 0.624, 0.333, 0.624) \label{w1machina}  \\
|w_2 \rangle&=&(0.342 e^{i 180^{\circ}}, 0.619 e^{i 270^{\circ}}, 0.342, 0.619 e^{i 90^{\circ}}) \nonumber \\
					&=&(-0.342, -0.619i, 0.342, 0.619i) \label{w2machina}
\end{eqnarray}
Also in this case, the solution is not unique.

The states $p_{w_1}$ and $p_{w_2}$ represented in (\ref{w1machina}) and (\ref{w2machina}) identify the subjective probability distributions $\mu_{w_1}$ and $\mu_{w_2}$, respectively, reproducing the experimental pattern. But, generally speaking, preferences do depend on the state $p_v$ of the DM entity. 

This completes the construction of a quantum model for the Machina reflection experiment with lower tail shifts, which follows the prescriptions in Section \ref{QEUT}.

`Machina reflection example with upper tail shifts'. The quantum-theoretic model is the same as the one constructed for the reflection example with lower tail shifts. The states $p_{w_1}$ and $p_{w_2}$ reproducing the preference weights 0.670 for $f_1$ over $f_2$ and 0.520 for $f_3$ over $f_4$ are respectively represented by the unit vectors
\begin{eqnarray}
|w_1 \rangle&=&(0.297, 0.642, 0.297, 0.642) \label{w1machina_upp}  \\
|w_2 \rangle&=&(0.353, 0.613 e^{i 90^{\circ}}, 0.353 e^{i 180^{\circ}}, 0.613 e^{i 270^{\circ}} ) \nonumber \\
					&=&(0.353, 0.613 i, -0.353 , -0.613 i ) \label{w2machina_upp}
\end{eqnarray}
for an utility value $u(50)-u(25) = \sqrt{50} - \sqrt{25} \approx 2.0711$, which completes the quantum representation of the Machina reflection experiment with upper tail shifts.


\section{Conclusions\label{conclusions}}
In this paper we tested the Ellsberg and Machina paradox situations in a decision-making experiment with human participants. The Ellsberg three-color experiment confirmed the typical ambiguity aversion pattern that is known in the literature, and the Machina 50/51 experiment confirmed the typical preferences suggested by Machina, against the predictions of SEUT and its major non-expected utility extensions. The Machina reflection experiments instead revealed a behavior that is not sensible to the informational symmetry suggested by Machina, which is compatible with the predictions of rank-dependent utility models, like Choquet expected utility.

We then showed that a quantum-theoretic framework developed by some of us \cite{ahs2017} enables modeling of various decision-making situations in presence of uncertainty, and allows faithful representation of the experiments above. This framework can be seen as the first step toward the development of a quantum-based state-dependent EUT.

We conclude this paper with some considerations on the `event-separability property' stated by SEUT through the sure-thing principle and partially retained by some of its extensions, like Choquet expected utility. According to Machina \cite{m2009}, the difficulties of SEUT to reproduce human decisions in Ellsberg-type scenarios stem from the event-separability property entailed by the sure-thing principle. Similarly, the difficulties of rank-dependent utility models, like Choquet expected utility, may be due to the `tail-separability property' entailed by its axiomatics. Indeed, the sure-thing principle states that preferences are separable across mutually exclusive events. A partial form of event-separability is retained by rank-dependent utility models, and formalized in a `comonotonic sure-thing principle' and the ensuing tail-separability. This does not instead occur in the quantum-theoretic framework, where preferences between acts are always connected through the state of the DM entity via (\ref{statedep}). The latter state also connects the mutually exclusive events, determining their subjective probabilities via (\ref{quantumprobability}). Hence, one cannot assume separability of acts across mutually exclusive events in the quantum-theoretic framework. This `non-separability property' might be responsible of the flexibility of the quantum-theoretic framework to accomodate human decisions in presence of ambiguity, in agreement with Machina's considerations.

\section{Acknowledgments}

This work was supported by Funda\c{c}\~{a}o para a Ci\^{e}ncia e a Tecnologia (FCT) with reference UID/CEC/50021/2013 and by the PhD grant SFRH/ BD/92391/2013. The funders had no role in study design, data collection and analysis, decision to publish, or preparation of the manuscript.

\appendix

\section{Essential quantum mathematics}\label{quantummathematics}
We present in this appendix the essential definitions and results of the mathematical formalism of quantum theory that are needed to represent DM entities and are useful to grasp the results attained in Section \ref{QEUT}. Our presentation will avoid superfluous technicalities, while aiming to be synthetic and rigorous.

When the quantum mechanical formalism is applied for modeling purposes, each considered entity  -- in our case a cognitive entity -- is associated with a complex Hilbert space ${\cal H}$, that is, a vector space over the field ${\mathbb C}$ of complex numbers, equipped with an inner product $\langle \cdot |  \cdot \rangle$ that maps two vectors $\langle A|$ and $|B\rangle$ onto a complex number $\langle A|B\rangle$. We denote vectors by using the bra-ket notation introduced by Paul Adrien Dirac, one of the pioneers of quantum theory \cite{d1958}. Vectors can be `kets', denoted by $\left| A \right\rangle $, $\left| B \right\rangle$, or `bras', denoted by $\left\langle A \right|$, $\left\langle B \right|$. The inner product between the ket vectors $|A\rangle$ and $|B\rangle$, or the bra-vectors $\langle A|$ and $\langle B|$, is realized by juxtaposing the bra vector $\langle A|$ and the ket vector $|B\rangle$, and $\langle A|B\rangle$ is also called a `bra-ket', and it satisfies the following properties:

(i) $\langle A |  A \rangle \ge 0$;

(ii) $\langle A |  B \rangle=\langle B |  A \rangle^{*}$, where $\langle B |  A \rangle^{*}$ is the complex conjugate of $\langle A |  B \rangle$;

(iii) $\langle A |(z|B\rangle+t|C\rangle)=z\langle A |  B \rangle+t \langle A |  C \rangle $, for $z, t \in {\mathbb C}$,
where the sum vector $z|B\rangle+t|C\rangle$ is called a `superposition' of vectors $|B\rangle$ and $|C\rangle$ in the quantum jargon.

From (ii) and (iii) follows that the inner product $\langle \cdot |  \cdot \rangle$ is linear in the ket and anti-linear in the bra, i.e. $(z\langle A|+t\langle B|)|C\rangle=z^{*}\langle A | C\rangle+t^{*}\langle B|C \rangle$.

We recall that the `absolute value' of a complex number is defined as the square root of the product of this complex number times its complex conjugate, that is, $|z|=\sqrt{z^{*}z}$. Moreover, a complex number $z$ can either be decomposed into its cartesian form $z=x+iy$, or into its polar form $z=|z|e^{i\theta}=|z|(\cos\theta+i\sin\theta)$.  As a consequence, we have $|\langle A| B\rangle|=\sqrt{\langle A|B\rangle\langle B|A\rangle}$. We define the `length' of a ket (bra) vector $|A\rangle$ ($\langle A|$) as $|| |A\rangle ||=||\langle A |||=\sqrt{\langle A |A\rangle}$. A vector of unitary length is called a `unit vector'. We say that the ket vectors $|A\rangle$ and $|B\rangle$ are `orthogonal' and write $|A\rangle \perp |B\rangle$ if $\langle A|B\rangle=0$.

We have now introduced the necessary mathematics to state the first modeling rule of quantum theory, as follows.

\medskip
\noindent{\it First quantum modeling rule:} A state $A$ of an entity -- in our case a cognitive entity -- modeled by quantum theory is represented by a ket vector $|A\rangle$ with length 1, that is $\langle A|A\rangle=1$.

\medskip
\noindent
An orthogonal projection $M$ is a linear operator on the Hilbert space, that is, a mapping $M: {\cal H} \rightarrow {\cal H}, |A\rangle \mapsto M|A\rangle$ which is Hermitian and idempotent. The latter means that, for every $|A\rangle, |B\rangle \in {\cal H}$ and $z, t \in {\mathbb C}$, we have:

(i) $M(z|A\rangle+t|B\rangle)=zM|A\rangle+tM|B\rangle$ (linearity);

(ii) $\langle A|M|B\rangle=\langle B|M|A\rangle^{*}$ (hermiticity);

(iii) $M \cdot M=M$ (idempotency).

The identity operator $\mathbbmss{1}$ maps each vector onto itself and is a trivial orthogonal projection operator. We say that two orthogonal projections $M_k$ and $M_l$ are orthogonal operators if each vector contained in the range $M_k({\cal H})$ is orthogonal to each vector contained in the range $M_l({\cal H})$, and we write $M_k \perp M_l$, in this case. The orthogonality of the projection operators $M_{k}$ and $M_{l}$ can also be expressed by $M_{k}M_{l}=0$, where $0$ is the null operator. A set of orthogonal projection operators $\{M_k \ | \ k\in \{1,\ldots,n \}\}$ is called a `spectral family' if all projectors are mutually orthogonal, that is, $M_k \perp M_l$ for $k \not= l$, and their sum is the identity, that is, $\sum_{k=1}^nM_k=\mathbbmss{1}$. A spectral family $\{M_k \ | \ k=1,\ldots,n\}$ identifies an Hermitian operator $\hat{O}=\sum_{i=1}^{n}o_kM_k$, where $o_k$ is called `eigenvalue of $\hat{O}$', i.e. is a solution of the equation $\hat{O}|o\rangle=o_k|o\rangle$ -- the non-null vectors satisfying this equation are called `eigenvectors of $\hat{O}$'.

The above definitions give us the necessary mathematics to state the second modeling rule of quantum theory, as follows.

\medskip
\noindent
{\it Second quantum modeling rule:} A measurable quantity $Q$ of an entity -- in our case a cognitive entity -- modeled by quantum theory, and having a set of possible real values $\{q_1, \ldots, q_n\}$ is represented by a spectral family $\{M_k \ | \ k=1, \ldots, n\}$, equivalently, by the Hermitian operator $\hat{Q}=\sum_{k=1}^{n}q_kM_k$, in the following way. If the conceptual entity is in a state represented by the vector $|A\rangle$, then the probability of obtaining the value $q_k$ in a measurement of the measurable quantity $Q$ is $\langle A|M_k|A\rangle=||M_k |A\rangle||^{2}$. This formula is called the `Born rule' in the quantum jargon. Moreover, if the value $q_k$ is actually obtained in the measurement, then the initial state is changed into a state represented by the vector
\begin{equation}
|A_k\rangle=\frac{M_k|A\rangle}{||M_k|A\rangle||}
\end{equation}
This change of state is called `collapse' in the quantum jargon.

\medskip
\noindent
Let us now come to the formalization of quantum probability. A major structural difference between classical probability theory, which satisfies the axioms of Kolmogorov, and quantum probability theory, which is non-Kolmo\-gorovian, relies on the fact that the former is defined on a Boolean $\sigma$-algebra of events, whilst the latter is defined on a more general algebraic structure \cite{p1989}. More specifically, let us denote by ${\mathscr L}({\cal H})$ the set of all orthogonal projection operators over the complex Hilbert space  ${\cal H}$. ${\mathscr L}({\cal H})$ has the algebraic properties of a complete orthocomplemented lattice, but ${\mathscr L}({\cal H})$  is not distributive, hence ${\mathscr L}({\cal H})$ does not form a $\sigma$-algebra. A `generalized probability measure' over  ${\mathscr L}({\cal H})$ is a function $\mu: M \in {\mathscr L}({\cal H}) \longmapsto \mu(M) \in [0,1]$, such that $\mu(\mathbbmss{1})=1$, and $\mu(\sum_{k=1}^{\infty}M_k)=\sum_{k=1}^{\infty}\mu(M_k)$, for any countable sequence $\{ M_k \in {\mathscr L}({\cal H}) \ | \  k\in \{1,2,\ldots\} \}$ of mutually orthogonal projection operators. The elements of ${\mathscr L}({\cal H})$ are said to represent `events', in this framework. Referring to the definitions above, the event ``a measurement of the quantity $Q$ gives the outcome $q_k$'' is represented by the orthogonal projection operator $M_k$.

Born's rule establishes a connection between states and generalized probability measures, as follows. 

Given a state of a cognitive entity represented by the vector $|A\rangle \in \cal H$ with length 1, it is possible to associate  $|A\rangle$ with a generalized probability measure $\mu_{A}$ over ${\mathscr L}({\cal H})$, such that, for every $M \in {\mathscr L}({\cal H})$, $\mu_{A}(M)=\langle A |M|A\rangle$. The generalized probability measure $\mu_{A}$ is a `quantum probability measure' over ${\mathscr L}({\cal H})$. Interestingly enough, if the dimension of the Hilbert space is greater than 3, all generalized probability measures over ${\mathscr L}({\cal H})$ can be written as functions $\mu_{A}(M)=\langle A|M|A\rangle$, for some unit vector $|A\rangle\in \cal H$. This is the content of the Gleason theorem \cite{g1957}.

The quantum theoretical modeling above can be extended by adding further quantum rules to model compound cognitive entities and more general classes of measurements on cognitive entities. However, the present definitions and results are sufficient to attain the results in this paper.










\end{document}